\documentclass[,final,numberedheadings]{aipproc}

\layoutstyle{6x9}

\usepackage{amsmath}
\usepackage{amsfonts}
\usepackage{amssymb}
\usepackage{graphicx}

\input{tcilatex}

\begin{document}

\title{Jacobi Fields on Statistical Manifolds of Negative Curvature}


\author{C. Cafaro\thanks{%
E-mail: carlocafaro2000@yahoo.it} \ and S. A. Ali\thanks{%
E-mail: alis@alum.rpi.edu}}{
address={Department of Physics, University at Albany--SUNY, Albany, NY 12222,USA}}

\begin{abstract}
Two entropic dynamical models are considered. The geometric structure of the
statistical manifolds underlying these models is studied. It is found that
in both cases, the resulting metric manifolds are negatively curved.
Moreover, the geodesics on each manifold are described by hyperbolic
trajectories. A detailed analysis based on the Jacobi equation for geodesic
spread is used to show that the hyperbolicity of the manifolds leads to
chaotic exponential instability. A comparison between the two models leads
to a relation among statistical curvature, stability of geodesics and
relative entropy-like quantities. Finally, the Jacobi vector field intensity
and the entropy-like quantity are suggested as possible indicators of
chaoticity in the ED models due to their similarity to the conventional
chaos indicators based on the Riemannian geometric approach and the
Zurek-Paz criterion of linear entropy growth, respectively.
\end{abstract}

\maketitle

\section{Introduction}

Entropic Dynamics (ED) $\left[ 1\right] $ is a theoretical framework
constructed on statistical manifolds to explore the possibility that laws of
physics, either classical or quantum, might be laws of inference rather than
laws of nature. It is known that thermodynamics can be obtained by means of
statistical mechanics which can be considered a form of statistical
inference $\left[ 2\right] $ rather than a pure physical theory. Indeed,
even some features of quantum physics can be derived from principles of
inference $\left[ 3\right] $. Finally, recent research considers the
possibility that Einstein's theory of gravity is derivable from general
principles of inductive inference $\left[ 4\right] $. Unfortunately, the
search for the correct variables that encode relevant information about a
system is a major obstacle in the description and understanding of its
evolution.\ The manner in which relevant variables are selected is not
straightforward. This selection is made, in most cases, on the basis of
intuition guided by experiment. The Maximum relative Entropy (ME) method $[5$%
, $6$, $7]$ is used to construct ED models. The ME method is designed to be
a tool of inductive inference. It is used for updating from a prior to a
posterior probability distribution when new information in the form of
constraints becomes available. We use known techniques $\left[ 1\right] $ to
show that this principle leads to equations that are analogous to equations
of motion. Information is processed using ME methods in the framework of
Information Geometry (IG) $\left[ 8\right] $ that is, Riemannian geometry
applied to probability theory. In our approach, probability theory is a form
of generalized logic of plausible inference. It should apply in principle,
to any situation where we lack sufficient information to permit deductive
reasoning.

In this paper, we focus on two special entropic dynamical models. In the
first model $\left( \text{ED1}\right) $, we consider an hypothetical system
whose microstates span a $2D$ space labelled by the variables $x_{1}\in 
\mathbb{R}
^{+}$ and $x_{2}\in 
\mathbb{R}
$. We assume that the only testable information pertaining to the quantities 
$x_{1}$ and $x_{2}$ consists of the expectation values $\left\langle
x_{1}\rangle \text{, }\langle x_{2}\right\rangle $ and the variance $\Delta
x_{2}$. In the second model $\left( \text{ED2}\right) $, we consider a $2D$
space of microstates labelled by the variables $x_{1}\in 
\mathbb{R}
$ and $x_{2}\in 
\mathbb{R}
$. In in this case, we assume that the only testable information pertaining
to the quantities $x_{1}$ and $x_{2}$ consist of the expectation values $%
\left\langle x_{1}\rangle \text{ and }\langle x_{2}\right\rangle $ and of
the variances $\Delta x_{1}$ and $\Delta x_{2}$. Our models may be extended
to more elaborate systems (highly constrained dynamics) where higher
dimensions are considered. However, for the sake of clarity, we restrict our
considerations to the above relatively simple cases. Given two known
boundary macrostates, we investigate the possible trajectories of systems on
the manifolds. The geometric structure of the manifolds underlying the
models is studied. The metric tensor, Christoffel connections coefficients,
Ricci and Riemann curvature tensors are calculated in both cases and it is
shown that in both cases the dynamics takes place on negatively curved
manifolds. The geodesics of the dynamical models are hyperbolic trajectories
on the manifolds. A detailed study of the stability of such geodesics is
presented using the equation of geodesic deviation (Jacobi equation). The
notion of statistical volume elements is introduced to investigate the
asymptotic behavior of a one-parameter family of neighboring geodesics. It
is shown that the behavior of geodesics on such manifolds is characterized
by exponential instability that leads to chaotic scenarios on the manifolds.
These conclusions are supported by the asymptotic behavior of the Jacobi
vector field intensity. Finally, a relation among entropy-like quantities,
instability and curvature in the two models is presented.

\section{Curved Statistical Manifolds}

In the case of ED1, a measure of distinguishability among the states of the
system is achieved by assigning a probability distribution $p\left( \vec{x}|%
\vec{\theta}\right) $ to each state defined by expected values $\theta
_{1}^{\left( 1\right) }$, $\theta _{1}^{\left( 2\right) }$, $\theta
_{2}^{\left( 2\right) }$ of the variables $x_{1}$, $x_{2}$ and $\left(
x_{2}-\left\langle x_{2}\right\rangle \right) ^{2}$. In the case of ED2, one
assigns a probability distribution $p\left( \vec{x}|\vec{\theta}\right) $ to
each state defined by expected values $\theta _{1}^{\left( 1\right) }$, $%
\theta _{2}^{\left( 1\right) }$, $\theta _{1}^{\left( 2\right) }$, $\theta
_{2}^{\left( 2\right) }$ \ of the variables $x_{1}$, $\left(
x_{1}-\left\langle x_{1}\right\rangle \right) ^{2}$, $x_{2}$ and $\left(
x_{2}-\left\langle x_{2}\right\rangle \right) ^{2}$. The process of
assigning a probability distribution to each state provides the statistical
manifolds of the ED models with a metric structure. Specifically, the
Fisher-Rao information metric $\left[ 9,10,11,12\right] $ defined in $(7)$
is used to quantify the distinguishability of probability distributions $%
p\left( \vec{x}|\vec{\theta}\right) $ that live on the manifold (the family
of distributions $\left\{ p^{(tot)}\left( \vec{x}|\vec{\theta}\right)
\right\} $ is as a manifold, each distribution $p^{(tot)}\left( \vec{x}|\vec{%
\theta}\right) $ is a point with coordinates $\theta ^{i}$ where $i$ labels
the macrovariables). As such, the Fisher-Rao metric assigns an IG to the
space of states.

\subsection{The Statistical Manifold $\mathcal{M}_{S_{1}}$}

Consider an hypothetical physical system evolving over a two-dimensional
space.\ The variables $x_{1}\in 
\mathbb{R}
^{+}$ and $x_{2}\in 
\mathbb{R}
$ label the $2D$ space of microstates of the system. We assume that all
information relevant to the dynamical evolution of the system is contained
in the probability distributions. For this reason, no other information
(such as external fields) is required. \ We assume that the only testable
information pertaining to the quantities $x_{1}$ and $x_{2}$ consists of the
expectation values $\left\langle x_{1}\rangle \text{, }\langle
x_{2}\right\rangle $ and the variance $\Delta x_{2}$. Therefore, these \
three expected values define the $3D$ space of macrostates $\mathcal{M}%
_{S_{1}}$ of the ED1 model. Each macrostate may be thought as a point of a
three-dimensional statistical manifold with coordinates given by the
numerical values of the expectations $\theta _{1}^{\left( 1\right) }$, $%
\theta _{1}^{\left( 2\right) }$, $\theta _{2}^{\left( 2\right) }$. The
available information can be written in the form of the following constraint
equations,%
\begin{equation}
\begin{array}{c}
\left\langle x_{1}\right\rangle =\int_{0}^{+\infty }dx_{1}x_{1}p_{1}\left(
x_{1}|\theta _{1}^{\left( 1\right) }\right) \text{, }\left\langle
x_{2}\right\rangle =\int_{-\infty }^{+\infty }dx_{2}x_{2}p_{2}\left(
x_{2}|\theta _{1}^{\left( 2\right) }\text{, }\theta _{2}^{\left( 2\right)
}\right) \text{,} \\ 
\\ 
\Delta x_{2}=\sqrt{\left\langle \left( x_{2}-\left\langle x_{2}\right\rangle
\right) ^{2}\right\rangle }=\left[ \int_{-\infty }^{+\infty }dx_{2}\left(
x_{2}-\left\langle x_{2}\right\rangle \right) ^{2}p_{2}\left( x_{2}|\theta
_{1}^{\left( 2\right) }\text{, }\theta _{2}^{\left( 2\right) }\right) \right]
^{\frac{1}{2}}\text{,}%
\end{array}%
\end{equation}%
where $\theta _{1}^{\left( 1\right) }=\left\langle x_{1}\right\rangle $, $%
\theta _{1}^{\left( 2\right) }=\left\langle x_{2}\right\rangle $ and $\theta
_{2}^{\left( 2\right) }=\Delta x_{2}$. The probability distributions $p_{1}$
and $p_{2}$ are constrained by the conditions of normalization,%
\begin{equation}
\int_{0}^{+\infty }dx_{1}p_{1}\left( x_{1}|\theta _{1}^{\left( 1\right)
}\right) =1\text{, }\int_{-\infty }^{+\infty }dx_{2}p_{2}\left( x_{2}|\theta
_{1}^{\left( 2\right) }\text{, }\theta _{2}^{\left( 2\right) }\right) =1%
\text{.}
\end{equation}%
Information theory identifies the exponential distribution as the maximum
entropy distribution if only the expectation value is known. The Gaussian
distribution is identified as the maximum entropy distribution if only the
expectation value and the variance are known. ME methods allow us to
associate a probability distribution $p^{(tot)}\left( \vec{x}|\vec{\theta}%
\right) $ to each point in the space of states. The distribution that best
reflects the information contained in the prior distribution $m\left( \vec{x}%
\right) $ updated by the constraints $(\left\langle x_{1}\right\rangle $, $%
\left\langle x_{2}\right\rangle $, $\Delta x_{2})$ is obtained by maximizing
the relative entropy%
\begin{equation}
\left[ S\left( \vec{\theta}\right) \right] _{\text{ED1}}=-\int_{0}^{+\infty
}\int_{-\infty }^{+\infty }dx_{1}dx_{2}p^{(tot)}\left( \overset{\rightarrow }%
{x}|\overset{\rightarrow }{\theta }\right) \log \left[ \frac{p^{(tot)}\left( 
\vec{x}|\vec{\theta}\right) }{m\left( \overset{\rightarrow }{x}\right) }%
\right] \text{,}
\end{equation}%
where $m(\vec{x})\equiv m$ is the uniform prior probability distribution.
The prior $m\left( \overset{\rightarrow }{x}\right) $ is set to be uniform
since we assume the lack of initial available information about the system
(postulate of equal \textit{a priori} probabilities). Upon maximizing $%
\left( 3\right) $, given the constraints $\left( 1\right) $ and $\left(
2\right) $, we obtain%
\begin{equation}
p^{(tot)}\left( \vec{x}|\vec{\theta}\right) =p_{1}\left( x_{1}|\theta
_{1}^{\left( 1\right) }\right) p_{2}\left( x_{2}|\theta _{1}^{\left(
2\right) }\text{, }\theta _{2}^{\left( 2\right) }\right) =\frac{1}{\mu _{1}}%
e^{-\frac{x_{1}}{\mu _{1}}}\frac{1}{\sqrt{2\pi \sigma _{2}^{2}}}e^{-\frac{%
(x_{2}-\mu _{2})^{2}}{2\sigma _{2}^{2}}}\text{,}
\end{equation}%
where $\theta _{1}^{\left( 1\right) }=\mu _{1}$, $\theta _{1}^{\left(
2\right) }=\mu _{2}$ and $\theta _{2}^{\left( 2\right) }=\sigma _{2}$. The
probability distribution $(4)$ encodes the available information concerning
the system and $\mathcal{M}_{s_{1}}$ becomes,%
\begin{equation}
\mathcal{M}_{s_{1}}=\left\{ p^{(tot)}\left( \vec{x}|\vec{\theta}\right) =%
\frac{1}{\mu _{1}}e^{-\frac{x_{1}}{\mu _{1}}}\frac{1}{\sqrt{2\pi \sigma
_{2}^{2}}}e^{-\frac{(x_{2}-\mu _{2})^{2}}{2\sigma _{2}^{2}}}\text{: }\vec{x}%
\in 
\mathbb{R}
^{+}\times 
\mathbb{R}
\text{ and }\vec{\theta}\equiv \left( \mu _{1}\text{, }\mu _{2}\text{, }%
\sigma _{2}\right) \right\} \text{.}
\end{equation}%
Note that we have assumed uncoupled constraints between the microvariables $%
x_{1}$ and $x_{2}$. In other words, we assumed that information about
correlations between the microvariables needed not to be tracked. This
assumption leads to the simplified product rule in $\left( 4\right) $.
Coupled constraints however, would lead to a generalized product rule in $%
\left( 4\right) $ and to a metric tensor $\left( 7\right) $ with non-trivial
off-diagonal elements (covariance terms). Correlation terms may be
fictitious. They may arise for instance from coordinate transformations. On
the other hand, correlations may arise from interaction of the system with
external fields. Such scenarios would require more delicate analysis.

\subsubsection{The Metric Tensor on $\mathcal{M}_{s_{1}}$}

We cannot determine the evolution of microstates of the system since the
available information is insufficient.\ Instead we can study the distance
between two total distributions with parameters $(\mu _{1}$, $\mu _{2}$, $%
\sigma _{2})$ and $(\mu _{1}+d\mu _{1}$, $\mu _{2}+d\mu _{2}$, $\sigma
_{2}+d\sigma _{2})$. Once the states of the system have been defined, the
next step concerns the problem of quantifying the notion of change in going
from the state $\vec{\theta}$ to the state $\vec{\theta}+d\vec{\theta}$. For
our purpose a convenient measure of change is distance. The measure we seek
is given by the dimensionless "distance" $ds$ between $p(\vec{x}|\vec{\theta}%
)$ and $p(\vec{x}|\vec{\theta}+d\vec{\theta})$:%
\begin{equation}
ds^{2}=g_{ij}d\theta ^{i}d\theta ^{j}\text{,}
\end{equation}%
where%
\begin{equation}
g_{ij}=\int d\vec{x}\text{ }p(\vec{x}|\vec{\theta})\frac{\partial \log p(%
\vec{x}|\vec{\theta})}{\partial \theta ^{i}}\frac{\partial \log p(\vec{x}|%
\vec{\theta})}{\partial \theta ^{j}}
\end{equation}%
is the Fisher-Rao information metric. Substituting $\left( 4\right) $ into $%
\left( 7\right) $, the metric $g_{ij}$ on $\mathcal{M}_{s_{1}}$ becomes,%
\begin{equation}
\left( g_{ij}\right) _{\mathcal{M}_{s_{1}}}=\left( 
\begin{array}{ccc}
\frac{1}{\mu _{1}^{2}} & 0 & 0 \\ 
0 & \frac{1}{\sigma _{2}^{2}} & 0 \\ 
0 & 0 & \frac{2}{\sigma _{2}^{2}}%
\end{array}%
\right) \text{.}
\end{equation}%
Substituting $\left( 8\right) $ into $\left( 6\right) $, the "length"
element reads,%
\begin{equation}
\left( ds^{2}\right) _{\mathcal{M}_{s_{1}}}=\frac{1}{\mu _{1}^{2}}d\mu
_{1}^{2}+\frac{1}{\sigma _{2}^{2}}d\mu _{2}^{2}+\frac{2}{\sigma _{2}^{2}}%
d\sigma _{2}^{2}\text{.}
\end{equation}%
Notice that the metric structure of $\mathcal{M}_{s_{1}}$ is an emergent
structure and is not itself fundamental. It arises only after assigning a
probability distribution $p\left( \vec{x}|\vec{\theta}\right) $ to each
state $\vec{\theta}$.

\subsubsection{The Curvature of $\mathcal{M}_{s_{1}}$}

In this paragraph we calculate the statistical curvature $R_{\mathcal{M}%
_{s_{1}}}$. This is achieved via application of standard differential
geometry methods to the space of probability distributions $\mathcal{M}%
_{s_{1}}$. Recall the definitions of the Ricci tensor $R_{ij}$ and Riemann
curvature tensor $R_{\alpha \mu \nu \rho }$, 
\begin{equation}
R_{ij}=g^{ab}R_{aibj}=\partial _{k}\Gamma _{ij}^{k}-\partial _{j}\Gamma
_{ik}^{k}+\Gamma _{ij}^{k}\Gamma _{kn}^{n}-\Gamma _{ik}^{m}\Gamma _{jm}^{k}%
\text{,}
\end{equation}%
and%
\begin{equation}
R_{\mu \nu \rho }^{\alpha }=\partial _{\nu }\Gamma _{\mu \rho }^{\alpha
}-\partial _{\rho }\Gamma _{\mu \nu }^{\alpha }+\Gamma _{\beta \nu }^{\alpha
}\Gamma _{\mu \rho }^{\beta }-\Gamma _{\beta \rho }^{\alpha }\Gamma _{\mu
\nu }^{\beta }\text{.}
\end{equation}%
The Ricci scalar $R_{\mathcal{M}_{s_{1}}}$ is obtained from $(10)$ or $%
\left( 11\right) $ via appropriate contraction with the metric tensor $g_{ij%
\text{ }}$in $\left( 8\right) $, namely%
\begin{equation}
R=R_{ij}g^{ij}=R_{\alpha \beta \gamma \delta }g^{\alpha \gamma }g^{\beta
\delta }\text{,}
\end{equation}%
where $g^{ik}g_{kj}=\delta _{j}^{i}$ so that $g^{ij}=\left( g_{ij}\right)
^{-1}=diag(\mu _{1}^{2}$, $\sigma _{2}^{2}$, $\frac{\sigma _{2}^{2}}{2})$.
The Christoffel symbols $\Gamma _{ij}^{k}$ appearing in $(10)$ and $\left(
11\right) $ are defined by,%
\begin{equation}
\Gamma _{ij}^{k}=\frac{1}{2}g^{km}\left( \partial _{i}g_{mj}+\partial
_{j}g_{im}-\partial _{m}g_{ij}\right) \text{.}
\end{equation}%
Substituting $(8)$ into $(13)$, we calculate the non-vanishing components of
the connection coefficients,%
\begin{equation}
\Gamma _{11}^{1}=-\frac{1}{\mu _{1}}\text{, }\Gamma _{22}^{3}=\frac{1}{%
2\sigma _{2}}\text{, }\Gamma _{33}^{3}=-\frac{1}{\sigma _{2}}\text{, }\Gamma
_{23}^{2}=\Gamma _{32}^{2}=-\frac{1}{\sigma _{2}}\text{.}
\end{equation}%
By substituting $(14)$ in $(10)$ we determine the Ricci tensor components,%
\begin{equation}
R_{11}=0\text{, }R_{22}=-\frac{1}{2\sigma _{2}^{2}}\text{, }R_{33}=-\frac{1}{%
\sigma _{2}^{2}}\text{.}
\end{equation}%
The non-vanishing Riemann tensor component is,%
\begin{equation}
R_{2323}=-\frac{1}{\sigma _{2}^{4}}\text{.}
\end{equation}%
Finally, by substituting $(15)$ or $\left( 16\right) $ into $(12)$ and using 
$\left( g_{ij}\right) ^{-1}$ we obtain the Ricci scalar,%
\begin{equation}
R_{\mathcal{M}_{s_{1}}}=-1<0\text{.}
\end{equation}%
From $(17)$ we conclude that $\mathcal{M}_{s_{1}}$ is a manifold of constant
negative $(-1)$ curvature.

\subsection{The Statistical Manifold $\mathcal{M}_{S_{2}}$}

In this case we assume that the $2D$ space of microstates of the system is
labelled by the variables $x_{1}\in 
\mathbb{R}
$ and $x_{2}\in 
\mathbb{R}
$. We assume, as in subsection $\left( 2.1\right) $, that all information
relevant to the dynamical evolution of the system is contained in the
probability distributions. Moreover, we assume that the only testable
information pertaining to the quantities $x_{1}$ and $x_{2}$ consist of the
expectation values $\left\langle x_{1}\rangle \text{ and }\langle
x_{2}\right\rangle $ and of the variances $\Delta x_{1}$ and $\Delta x_{2}$.
Therefore, these four expected values define the $4D$ space of macrostates $%
\mathcal{M}_{S_{2}}$ of the ED2 model. Each macrostate may be thought as a
point of a four-dimensional statistical manifold with coordinates given by
the numerical values of the expectations $\theta _{1}^{\left( 1\right) }$, $%
\theta _{2}^{\left( 1\right) }$, $\theta _{1}^{\left( 2\right) }$, $\theta
_{2}^{\left( 2\right) }$. The available information can be written in the
form of the following constraint equations,%
\begin{equation}
\begin{array}{c}
\left\langle x_{1}\right\rangle =\int_{-\infty }^{+\infty
}dx_{1}x_{1}p_{1}\left( x_{1}|\theta _{1}^{\left( 1\right) }\right) \text{,
\ }\left\langle x_{2}\right\rangle =\int_{-\infty }^{+\infty
}dx_{2}x_{2}p_{2}\left( x_{2}|\theta _{1}^{\left( 2\right) }\text{, }\theta
_{2}^{\left( 2\right) }\right) \text{,} \\ 
\\ 
\Delta x_{1}=\sqrt{\left\langle \left( x_{1}-\left\langle x_{1}\right\rangle
\right) ^{2}\right\rangle }=\left[ \int_{-\infty }^{+\infty }dx_{1}\left(
x_{1}-\left\langle x_{1}\right\rangle \right) ^{2}p_{1}\left( x_{1}|\theta
_{1}^{\left( 1\right) }\text{, }\theta _{2}^{\left( 1\right) }\right) \right]
^{\frac{1}{2}}\text{,} \\ 
\\ 
\Delta x_{2}=\sqrt{\left\langle \left( x_{2}-\left\langle x_{2}\right\rangle
\right) ^{2}\right\rangle }=\left[ \int_{-\infty }^{+\infty }dx_{2}\left(
x_{2}-\left\langle x_{2}\right\rangle \right) ^{2}p_{2}\left( x_{2}|\theta
_{1}^{\left( 2\right) }\text{, }\theta _{2}^{\left( 2\right) }\right) \right]
^{\frac{1}{2}}\text{,}%
\end{array}%
\end{equation}%
where $\theta _{1}^{\left( 1\right) }=\left\langle x_{1}\right\rangle $, $%
\theta _{2}^{\left( 1\right) }=\Delta x_{1}$, $\theta _{1}^{\left( 2\right)
}=\left\langle x_{2}\right\rangle $ and $\theta _{2}^{\left( 2\right)
}=\Delta x_{2}$. The probability distributions $p_{1}$ and $p_{2}$ are
constrained by the conditions of normalization,%
\begin{equation}
\int_{-\infty }^{+\infty }dx_{1}p_{1}\left( x_{1}|\theta _{1}^{\left(
1\right) }\text{, }\theta _{2}^{\left( 1\right) }\right) =1\text{, }%
\int_{-\infty }^{+\infty }dx_{2}p_{2}\left( x_{2}|\theta _{1}^{\left(
2\right) }\text{, }\theta _{2}^{\left( 2\right) }\right) =1\text{.}
\end{equation}%
The distribution that best reflects the information contained in the uniform
prior distribution $m\left( \vec{x}\right) \equiv m$ updated by the
constraints $(\left\langle x_{1}\right\rangle $, $\Delta x_{1}$, $%
\left\langle x_{2}\right\rangle $, $\Delta x_{2})$ is obtained by maximizing
the relative entropy%
\begin{equation}
\left[ S\left( \vec{\theta}\right) \right] _{\text{ED2}}=-\int_{-\infty
}^{+\infty }\int_{-\infty }^{+\infty }dx_{1}dx_{2}p^{(tot)}\left( \overset{%
\rightarrow }{x}|\overset{\rightarrow }{\theta }\right) \log \left[ \frac{%
p^{(tot)}\left( \vec{x}|\vec{\theta}\right) }{m\left( \overset{\rightarrow }{%
x}\right) }\right] \text{.}
\end{equation}%
Upon maximizing $\left( 20\right) $, given the constraints $\left( 18\right) 
$ and $\left( 19\right) $, we obtain%
\begin{equation}
p^{(tot)}\left( \vec{x}|\vec{\theta}\right) =\frac{1}{\sqrt{2\pi \sigma
_{1}^{2}}}e^{-\frac{(x_{1}-\mu _{1})^{2}}{2\sigma _{1}^{2}}}\frac{1}{\sqrt{%
2\pi \sigma _{2}^{2}}}e^{-\frac{(x_{2}-\mu _{2})^{2}}{2\sigma _{2}^{2}}}%
\text{.}
\end{equation}%
The probability distribution $(21)$ encodes the available information
concerning the system and $\mathcal{M}_{s_{2}}$ becomes,%
\begin{equation}
\mathcal{M}_{s_{2}}=\left\{ p^{(tot)}\left( \vec{x}|\vec{\theta}\right) =%
\frac{1}{\sqrt{2\pi \sigma _{1}^{2}}}e^{-\frac{(x_{1}-\mu _{1})^{2}}{2\sigma
_{1}^{2}}}\frac{1}{\sqrt{2\pi \sigma _{2}^{2}}}e^{-\frac{(x_{2}-\mu _{2})^{2}%
}{2\sigma _{2}^{2}}}\text{: }\vec{x}\in 
\mathbb{R}
\times 
\mathbb{R}
\text{ and }\vec{\theta}\equiv \left( \mu _{1}\text{, }\sigma _{1}\text{, }%
\mu _{2}\text{, }\sigma _{2}\right) \right\} \text{.}
\end{equation}

\subsubsection{The Metric Tensor on $\mathcal{M}_{s_{2}}$}

Proceeding as in $\left( 2.1.1\right) $, we determine the metric on $%
\mathcal{M}_{s_{2}}$. Substituting $\left( 21\right) $ into $\left( 7\right) 
$, the metric $g_{ij}$ on $\mathcal{M}_{s_{2}}$ becomes,%
\begin{equation}
\left( g_{ij}\right) _{\mathcal{M}_{s_{2}}}=\left( 
\begin{array}{cccc}
\frac{1}{\sigma _{1}^{2}} & 0 & 0 & 0 \\ 
0 & \frac{2}{\sigma _{1}^{2}} & 0 & 0 \\ 
0 & 0 & \frac{1}{\sigma _{2}^{2}} & 0 \\ 
0 & 0 & 0 & \frac{2}{\sigma _{2}^{2}}%
\end{array}%
\right) \text{.}
\end{equation}%
Finally, substituting $\left( 23\right) $ into $\left( 6\right) $, the
"length" element reads,%
\begin{equation}
\left( ds^{2}\right) _{\mathcal{M}_{s_{2}}}=\frac{1}{\sigma _{1}^{2}}d\mu
_{1}^{2}+\frac{2}{\sigma _{1}^{2}}d\sigma _{1}^{2}+\frac{1}{\sigma _{2}^{2}}%
d\mu _{2}^{2}+\frac{2}{\sigma _{2}^{2}}d\sigma _{2}^{2}\text{.}
\end{equation}

\subsubsection{The Curvature of $\mathcal{M}_{s_{2}}$}

Proceeding as in $\left( 2.1.2\right) $, we calculate the statistical
curvature $R_{\mathcal{M}_{s_{2}}}$ of $\mathcal{M}_{s_{2}}$. Notice that $%
g^{ij}=\left( g_{ij}\right) ^{-1}=diag(\sigma _{1}^{2}$, $\frac{\sigma
_{1}^{2}}{2}$, $\sigma _{2}^{2}$, $\frac{\sigma _{2}^{2}}{2})$. Substituting 
$(23)$ into $(13)$, the non-vanishing components of the connection
coefficients become,%
\begin{equation}
\Gamma _{12}^{1}=\Gamma _{21}^{1}=-\frac{1}{\sigma _{1}}\text{, }\Gamma
_{22}^{2}=-\frac{1}{\sigma _{1}}\text{, }\Gamma _{11}^{2}=\frac{1}{2\sigma
_{1}}\text{, }\Gamma _{34}^{3}=\Gamma _{43}^{3}=-\frac{1}{\sigma _{2}}\text{%
, }\Gamma _{33}^{4}=\frac{1}{2\sigma _{2}}\text{, }\Gamma _{44}^{4}=-\frac{1%
}{\sigma _{2}}\text{.}
\end{equation}%
By substituting $(25)$ in $(10)$ we determine the Ricci tensor components,%
\begin{equation}
R_{11}=-\frac{1}{2\sigma _{1}^{2}}\text{, }R_{22}=-\frac{1}{\sigma _{1}^{2}}%
\text{, }R_{33}=-\frac{1}{2\sigma _{2}^{2}}\text{, }R_{44}=-\frac{1}{\sigma
_{2}^{2}}\text{.}
\end{equation}%
The non-vanishing Riemann tensor components are,%
\begin{equation}
R_{1212}=-\frac{1}{\sigma _{1}^{4}}\text{, }R_{3434}=-\frac{1}{\sigma
_{2}^{4}}\text{.}
\end{equation}%
Finally, by substituting $(26)$ or $\left( 27\right) $ into $(12)$ and using 
$\left( g_{ij}\right) ^{-1}$, we obtain the Ricci scalar,%
\begin{equation}
R_{\mathcal{M}_{s_{2}}}=-2<0\text{.}
\end{equation}%
From $(28)$ we conclude that $\mathcal{M}_{s_{2}}$ is a manifold of constant
negative $(-2)$ curvature.

\section{The ED\ Models}

The ED models can be derived from a standard principle of least action
(Maupertuis- Euler-Lagrange-Jacobi-type) $[1$, $13]$. The main differences
are that the dynamics being considered here is defined on a space of
probability distributions $\mathcal{M}_{s}$, not on an ordinary linear space 
$V$. Also, the standard coordinates $q_{j}$ of the system are replaced by
statistical macrovariables $\theta ^{j}$.

Given the initial macrostate and that the system evolves to a fixed final
macrostate, we investigate the expected trajectories of the ED models on $%
\mathcal{M}_{s_{1}}$ and $\mathcal{M}_{s_{2}}$. It is known $\left[ 13\right]
$ that the classical dynamics of a particle can be derived from the
principle of least action in the Maupertuis-Euler-Lagrange-Jacobi form,%
\begin{equation}
\delta J_{\text{Jacobi}}\left[ q\right] =\delta \int_{s_{i}}^{s_{f}}ds%
\mathcal{F}\left( q_{j}\text{, }\frac{dq_{j}}{ds}\text{, }s\text{, }H\right)
=0\text{,}
\end{equation}%
where $q_{j}$ are the coordinates of the system, $s$ is an affine parameter
along the trajectory and $\mathcal{F}$ \ is a functional defined as%
\begin{equation}
\mathcal{F}\left( q_{j}\text{, }\frac{dq_{j}}{ds}\text{, }s\text{, }H\right)
\equiv \left[ 2\left( H-U\right) \right] ^{\frac{1}{2}}\left( \underset{j,k}{%
\sum }a_{jk}\frac{dq_{j}}{ds}\frac{dq_{k}}{ds}\right) ^{\frac{1}{2}}\text{.}
\end{equation}%
For a non-relativistic system, the energy $H$ is,%
\begin{equation}
H=T+U\left( q\right) =\frac{1}{2}\underset{j,k}{\sum }a_{jk}\left( q\right) 
\overset{\cdot }{q}_{j}\overset{\cdot }{q}_{k}+U\left( q\right)
\end{equation}%
where the coefficients $a_{jk}$ are the reduced mass matrix coefficients and 
$\overset{\cdot }{q}=\frac{dq}{ds}$ is the time derivative of the canonical
coordinate $q$. We now seek the expected trajectory of the system assuming
it evolves from $\theta _{old}^{\mu }=\theta ^{\mu }$\ $\equiv \left( \mu
_{1}\left( s_{i}\right) \text{, }\mu _{2}\left( s_{i}\right) \text{, }\sigma
_{2}\left( s_{i}\right) \right) $ to $\theta _{new}^{\mu }=\theta ^{\mu
}+d\theta ^{\mu }\equiv \left( \mu _{1}\left( s_{f}\right) \text{, }\mu
_{2}\left( s_{f}\right) \text{, }\sigma _{2}\left( s_{f}\right) \right) $.
It is known $\left[ 1\right] $ that such a system moves along a geodesic in
the space of states, which is a curved manifold.\ Since the trajectory of
the system is a geodesic, the ED-action is itself the length; that is,%
\begin{equation}
J_{ED}\left[ \theta \right] =\int \left( ds^{2}\right) ^{\frac{1}{2}}=\int
\left( g_{ij}d\theta ^{i}d\theta ^{j}\right) ^{\frac{1}{2}%
}=\int_{s_{i}}^{s_{f}}ds\left( g_{ij}\frac{d\theta ^{i}\left( s\right) }{ds}%
\frac{d\theta ^{j}\left( s\right) }{ds}\right) ^{\frac{1}{2}}\equiv
\int_{s_{i}}^{s_{f}}ds\mathcal{L}(\theta ,\overset{\cdot }{\theta })
\end{equation}%
where $\overset{\cdot }{\theta }=\frac{d\theta }{ds}$ and $\mathcal{L}%
(\theta ,\overset{\cdot }{\theta })$ is the Lagrangian of the system,%
\begin{equation}
\mathcal{L}(\theta \text{,}\overset{\cdot }{\theta })=(g_{ij}\overset{\cdot }%
{\theta ^{i}}\overset{\cdot }{\theta ^{j}})^{\frac{1}{2}}\text{.}
\end{equation}%
A useful choice for $s$ is one satisfying the condition $g_{ij}\frac{d\theta
^{i}}{ds}\frac{d\theta ^{j}}{ds}=1$. Therefore, we formally identify the
affine parameter $s$ with the temporal evolution parameter $\tau $, $s\equiv
\tau $. Performing a standard calculus of variations with $s\equiv \tau $,
we obtain%
\begin{equation}
\delta J_{ED}\left[ \theta \right] =\int d\tau \left( \frac{1}{2}\frac{%
\partial g_{ij}}{\partial \theta ^{k}}\overset{\cdot }{\theta ^{i}}\overset{%
\cdot }{\theta ^{j}}-\frac{d\overset{\cdot }{\theta }_{k}}{d\tau }\right)
\delta \theta ^{k}=0\text{, }\forall \delta \theta ^{k}\text{.}
\end{equation}%
Note that from $(34)$, $\frac{d\overset{\cdot }{\theta }_{k}}{d\tau }=$ $%
\frac{1}{2}\frac{\partial g_{ij}}{\partial \theta ^{k}}\overset{\cdot }{%
\theta ^{i}}\overset{\cdot }{\theta ^{j}}$. This differential equation shows
that if $\frac{\partial g_{ij}}{\partial \theta ^{k}}=0$ for a particular $k$
then the corresponding $\overset{\cdot }{\theta }_{k}$ is conserved. This
suggests to interpret $\overset{\cdot }{\theta }_{k}$ as momenta. Equations $%
\left( 34\right) $ and $\left( 13\right) $ lead to the geodesic equation,%
\begin{equation}
\frac{d^{2}\theta ^{k}(\tau )}{d\tau ^{2}}+\Gamma _{ij}^{k}\frac{d\theta
^{i}(\tau )}{d\tau }\frac{d\theta ^{j}(\tau )}{d\tau }=0\text{.}
\end{equation}%
Observe that $\left( 35\right) $ are \textit{nonlinear}, second order
coupled ordinary differential equations. These equations describe a dynamics
that is reversible and their solution is the trajectory between an initial
and a final macrostate. The trajectory can be equally well traversed in both
directions.

\subsection{The ED1: Geodesics on $\mathcal{M}_{s_{1}}$}

We seek the explicit form of $(35)$ for ED1. Substituting $\left( 14\right) $
in $\left( 35\right) $, we obtain,%
\begin{equation}
\frac{d^{2}\mu _{1}}{d\tau ^{2}}=\frac{1}{\mu _{1}}\left( \frac{d\mu _{1}}{%
d\tau }\right) ^{2}\text{, \ }\frac{d^{2}\mu _{2}}{d\tau ^{2}}=\frac{2}{%
\sigma _{2}}\frac{d\mu _{2}}{d\tau }\frac{d\sigma _{2}}{d\tau }\text{, }%
\frac{d^{2}\sigma _{2}}{d\tau ^{2}}=\frac{1}{\sigma _{2}}\left[ \left( \frac{%
d\sigma _{2}}{d\tau }\right) ^{2}-\frac{1}{2}\left( \frac{d\mu _{2}}{d\tau }%
\right) ^{2}\right] \text{.}
\end{equation}%
Integrating this set of differential equations, we obtain 
\begin{equation}
\begin{array}{c}
\mu _{1}\left( \tau \right) =A_{1}\left[ \cosh \left( \alpha _{1}\tau
\right) -\sinh \left( \alpha _{1}\tau \right) \right] \text{,} \\ 
\\ 
\mu _{2}\left( \tau \right) =\frac{B_{1}^{2}}{2\beta _{1}}\frac{1}{\cosh
\left( 2\beta _{1}\tau \right) -\sinh \left( 2\beta _{1}\tau \right) +\frac{%
B_{1}^{2}}{8\beta _{1}^{2}}}+C_{1}\text{, \ }\sigma _{2}\left( \tau \right)
=B_{1}\frac{\left[ \cosh \left( \beta _{1}\tau \right) -\sinh \left( \beta
_{1}\tau \right) \right] }{\cosh \left( 2\beta _{1}\tau \right) -\sinh
\left( 2\beta _{1}\tau \right) +\frac{B_{1}^{2}}{8\beta _{1}^{2}}}\text{,}%
\end{array}%
\end{equation}%
where $A_{1}$, $B_{1}$, $C_{1}$, $\alpha _{1}$ and $\beta _{1}$ are \textit{%
real} integration constants ($5=6-1$, $\left( \dot{\theta}_{j}\dot{\theta}%
^{j}\right) ^{\frac{1}{2}}=1$) . The set of equations $(37)$ parametrizes
the evolution surface of \ the statistical submanifold $\mathit{m}_{s_{1}}$
of $\mathcal{M}_{s_{1}}$,%
\begin{equation}
\mathit{m}_{_{s_{1}}}=\left\{ p^{(tot)}\left( \vec{x}|\vec{\theta}\right)
\in \mathcal{M}_{s_{1}}\text{: }\vec{\theta}\text{ satisfy }\left( 36\right)
\right\} \text{.}
\end{equation}

\subsection{The ED2: Geodesics on $\mathcal{M}_{s_{2}}$}

We seek the explicit form of $(35)$ for ED2. Substituting $\left( 25\right) $
in $\left( 35\right) $, we obtain,%
\begin{equation}
\begin{array}{c}
\frac{d^{2}\mu _{1}}{d\tau ^{2}}=\frac{2}{\sigma _{1}}\frac{d\mu _{1}}{d\tau 
}\frac{d\sigma _{1}}{d\tau }\text{, }\frac{d^{2}\sigma _{1}}{d\tau ^{2}}=%
\frac{1}{\sigma _{1}}\left[ \left( \frac{d\sigma _{1}}{d\tau }\right) ^{2}-%
\frac{1}{2}\left( \frac{d\mu _{1}}{d\tau }\right) ^{2}\right] \text{,} \\ 
\\ 
\frac{d^{2}\mu _{2}}{d\tau ^{2}}=\frac{2}{\sigma _{2}}\frac{d\mu _{2}}{d\tau 
}\frac{d\sigma _{2}}{d\tau }\text{, }\frac{d^{2}\sigma _{2}}{d\tau ^{2}}=%
\frac{1}{\sigma _{2}}\left[ \left( \frac{d\sigma _{2}}{d\tau }\right) ^{2}-%
\frac{1}{2}\left( \frac{d\mu _{2}}{d\tau }\right) ^{2}\right] \text{.}%
\end{array}%
\end{equation}%
Integrating this set of differential equations, we obtain%
\begin{equation}
\begin{array}{c}
\mu _{1}\left( \tau \right) =\frac{A_{2}^{2}}{2\alpha _{2}}\frac{1}{\cosh
\left( 2\alpha _{2}\tau \right) -\sinh \left( 2\alpha _{2}\tau \right) +%
\frac{A_{2}^{2}}{8\alpha _{2}^{2}}}+C_{1}\text{, \ }\sigma _{1}\left( \tau
\right) =A_{2}\frac{\left[ \cosh \left( \alpha _{2}\tau \right) -\sinh
\left( \alpha _{2}\tau \right) \right] }{\cosh \left( 2\alpha _{2}\tau
\right) -\sinh \left( 2\alpha _{2}\tau \right) +\frac{A_{2}^{2}}{8\alpha
_{2}^{2}}}\text{,} \\ 
\\ 
\mu _{2}\left( \tau \right) =\frac{B_{2}^{2}}{2\beta _{2}}\frac{1}{\cosh
\left( 2\beta _{2}\tau \right) -\sinh \left( 2\beta _{2}\tau \right) +\frac{%
B_{2}^{2}}{8\beta _{2}^{2}}}+C_{2}\text{, }\sigma _{2}\left( \tau \right)
=B_{2}\frac{\left[ \cosh \left( \beta _{2}\tau \right) -\sinh \left( \beta
_{2}\tau \right) \right] }{\cosh \left( 2\beta _{2}\tau \right) -\sinh
\left( 2\beta _{2}\tau \right) +\frac{B_{2}^{2}}{8\beta _{2}^{2}}}\text{,}%
\end{array}%
\end{equation}%
where $A_{2}$, $B_{2}$, $C_{1}$, $C_{2}$, $\alpha _{2}$ and $\beta _{2}$ are 
\textit{real} integration constants. The set of equations $(40)$
parametrizes the evolution surface of \ the statistical submanifold $\mathit{%
m}_{s_{2}}$ of $\mathcal{M}_{s_{2}}$,%
\begin{equation}
\mathit{m}_{_{s_{2}}}=\left\{ p^{(tot)}\left( \vec{x}|\vec{\theta}\right)
\in \mathcal{M}_{s_{2}}\text{: }\vec{\theta}\text{ satisfy }\left( 39\right)
\right\} \text{.}
\end{equation}

\section{Chaotic Instability in the ED Models}

It is known that $\left[ 13\right] $ the Riemannian curvature of a manifold
is closely connected with the behavior of the geodesics on it, i.e., with
the motion of the corresponding dynamical system. If the Riemannian
curvature of a manifold is positive (as on a sphere or ellipsoid), then the
nearby geodesics oscillate about one another in most cases; whereas if the
curvature is negative (as on the surface of a hyperboloid of one sheet),
geodesics rapidly diverge from one another.

\subsection{Instability in ED1}

In this subsection, the stability of ED1 is considered. It is shown that
neighboring trajectories are exponentially unstable under small
perturbations of initial conditions. In the rest of the paper, for the sake
of simplicity, we assume that $A_{1}=B_{1}=A_{2}=B_{2}\equiv A$, $%
C_{1}=C_{2}=0$ and $\alpha _{1}=\beta _{1}=\alpha _{2}=\beta _{2}\equiv
\alpha $. Our conclusions do not depend on the particular initial conditions
chosen.

\subsubsection{The Geodesic Length $\Theta _{\mathcal{M}_{s_{1}}}$}

Consider the one-parameter family of geodesics $\mathcal{F}_{G_{\mathcal{M}%
_{s_{1}}}}\left( \alpha \right) \equiv \left\{ \theta _{\mathcal{M}%
_{s_{1}}}^{\mu }\left( \tau ;\alpha \right) \right\} _{\alpha \in 
\mathbb{R}
^{+}}^{\mu =1\text{, }2\text{, }3}$ where $\theta _{\mathcal{M}%
_{s_{1}}}^{\mu }$are solutions of $\left( 36\right) $. The length of
geodesics in $\mathcal{F}_{G_{\mathcal{M}_{s_{1}}}}\left( \alpha \right) $
is defined as,%
\begin{equation}
\Theta _{\mathcal{M}_{s_{1}}}\left( \tau \text{; }\alpha \right) \overset{%
\text{def}}{=}\dint \left( g_{ij}d\theta ^{i}d\theta ^{j}\right) ^{\frac{1}{2%
}}=\dint\limits_{0}^{\tau }\left[ \frac{1}{\mu _{1}^{2}}\left( \frac{d\mu
_{1}}{d\tau ^{\prime }}\right) ^{2}+\frac{1}{\sigma _{2}^{2}}\left( \frac{%
d\mu _{2}}{d\tau ^{\prime }}\right) ^{2}+\frac{2}{\sigma _{2}^{2}}\left( 
\frac{d\sigma _{2}}{d\tau ^{\prime }}\right) ^{2}\right] ^{\frac{1}{2}}d\tau
^{\prime }\text{.}
\end{equation}%
Substituting $\left( 37\right) $ in $\left( 42\right) $ and considering the
asymptotic expression of $\Theta _{\mathcal{M}_{s_{1}}}\left( \tau \text{; }%
\alpha \right) $, we obtain%
\begin{equation}
\Theta _{\mathcal{M}_{s_{1}}}\left( \tau \rightarrow \infty \text{; }\alpha
\right) \equiv \Theta _{1}\left( \tau ;\alpha \right) \approx \sqrt{3}\alpha
\tau \text{.}
\end{equation}%
In order to investigate the asymptotic behavior of two neighboring geodesics
labelled by the parameters $\alpha $ and $\alpha +\delta \alpha $, we
consider the following difference,%
\begin{equation}
\Delta \Theta _{1}\equiv \left\vert \Theta _{1}\left( \tau \text{; }\alpha
+\delta \alpha \right) -\Theta _{1}\left( \tau \text{; }\alpha \right)
\right\vert =\sqrt{3}\left\vert \delta \alpha \right\vert \tau \text{.}
\end{equation}%
It is clear that $\Delta \Theta _{1}$ diverges, that is, the lengths of two
neighboring geodesics\ with slightly different parameters $\alpha $ and $%
\alpha +\delta \alpha $ differ in a remarkable way as the evolution
parameter $\tau $ $\rightarrow \infty $. This hints at the onset of
instability of the hyperbolic trajectories on $\mathcal{M}_{s_{1}}$.

\subsubsection{The Statistical Volume Elements $V_{\mathcal{M}_{s_{1}}}$}

The instability of ED1 can be further explored by studying the behavior of
the one-parameter family of statistical volume elements $\mathcal{F}_{V_{%
\mathcal{M}_{s_{1}}}}\left( \alpha \right) \equiv \left\{ V_{\mathcal{M}%
_{s_{1}}}\left( \tau \text{; }\alpha \right) \right\} _{\alpha }$. Recall
that $\mathcal{M}_{s_{1}}$ is the space of probability distributions $%
p^{(tot)}\left( \vec{x}|\vec{\theta}\right) $ labeled by parameters $\theta
_{1}^{\left( 1\right) }$, $\theta _{1}^{\left( 2\right) }$, $\theta
_{2}^{\left( 2\right) }$. These parameters are the coordinates of the point $%
p^{(tot)}$, and in these coordinates a $3D$ volume element $dV_{\mathcal{M}%
_{s_{1}}}$ reads%
\begin{equation}
dV_{\mathcal{M}_{s_{1}}}=\sqrt{g}d\theta _{1}^{\left( 1\right) }d\theta
_{1}^{\left( 2\right) }d\theta _{2}^{\left( 2\right) }\equiv \sqrt{g}d\mu
_{1}d\mu _{2}d\sigma _{2}\text{,}
\end{equation}%
where in the ED1 model here presented, $g=|\det \left( g_{ij}\right) _{%
\mathcal{M}_{s_{1}}}|=\frac{2}{\mu _{1}^{2}\sigma _{2}^{4}}$. Hence, the
volume element $dV_{\mathcal{M}_{s_{1}}}$ is given by,%
\begin{equation}
dV_{\mathcal{M}_{s_{1}}}=\frac{\sqrt{2}}{\mu _{1}\sigma _{2}^{2}}d\mu
_{1}d\mu _{2}d\sigma _{2}\text{.}
\end{equation}%
The volume of an extended region of $\mathcal{M}_{s_{1}}$ is defined by,%
\begin{equation}
\Delta V_{\mathcal{M}_{s_{1}}}\equiv V_{\mathcal{M}_{s_{1}}}\left( \tau
\right) -V_{\mathcal{M}_{s_{1}}}\left( 0\right) \overset{\text{def}}{=}%
\dint\limits_{V_{\mathcal{M}_{s_{1}}}\left( 0\right) }^{V_{\mathcal{M}%
_{s_{1}}}\left( \tau \right) }dV_{\mathcal{M}_{s_{1}}}=\dint\limits_{\mu
_{1}\left( 0\right) }^{\mu _{1}\left( \tau \right) }\dint\limits_{\mu
_{2}\left( 0\right) }^{\mu _{2}\left( \tau \right) }\dint\limits_{\sigma
_{2}\left( 0\right) }^{\sigma _{2}\left( \tau \right) }\frac{\sqrt{2}}{\mu
_{1}\sigma _{2}^{2}}d\mu _{1}d\mu _{2}d\sigma _{2}\text{.}
\end{equation}%
Integrating $\left( 47\right) $ using $\left( 37\right) $, we obtain%
\begin{equation}
\Delta V_{\mathcal{M}_{s_{1}}}=\frac{\tau }{\sqrt{2}}e^{\alpha \tau }-\frac{%
\ln A}{\sqrt{2}\alpha }e^{\alpha \tau }+\frac{\ln A}{\sqrt{2}\alpha }\text{.}
\end{equation}%
The quantity that actually encodes relevant information about the stability
of neighboring volume elements is the the average volume $\left\langle
\Delta V_{\mathcal{M}_{s_{1}}}\right\rangle _{\tau }$, 
\begin{equation}
\left\langle \Delta V_{\mathcal{M}_{s_{1}}}\right\rangle _{\tau }\overset{%
\text{def}}{=}\frac{1}{\tau }\dint\limits_{0}^{\tau }\Delta V_{\mathcal{M}%
_{s_{1}}}\left( \tau ^{\prime }\text{; }\alpha \right) d\tau ^{\prime }=%
\frac{1}{\tau }\left\{ \frac{1}{\sqrt{2}\alpha ^{2}}\left( \alpha \tau
-1\right) e^{\alpha \tau }-\frac{\ln A}{\sqrt{2}\alpha ^{2}}e^{\alpha \tau }+%
\frac{\ln A}{\sqrt{2}\alpha }\tau \right\} \text{.}
\end{equation}%
For convenience, let us rename $\left\langle \Delta V_{\mathcal{M}%
_{s_{1}}}\right\rangle _{\tau }\equiv \Delta V_{1}$. Therefore, the
asymptotic expansion of $\Delta V_{1}$ for $\tau $ $\rightarrow \infty $
reads,%
\begin{equation}
\Delta V_{1}\approx \frac{1}{\sqrt{2}\alpha }e^{\alpha \tau }\text{.}
\end{equation}%
This asymptotic regime of diffusive evolution in $\left( 50\right) $
describes the exponential increase of average volume elements on $\mathcal{M}%
_{s_{1}}$. The exponential instability characteristic of chaos forces the
system to rapidly explore large areas (volumes) of the statistical
manifolds. It is interesting to note that this asymptotic behavior appears
also in the conventional description of quantum chaos $\left[ 14\right] $
where the entropy increases linearly at a rate determined by the Lyapunov
exponents $\left[ 15\right] $. The linear entropy increase as a quantum
chaos criterion was introduced by Zurek and Paz. In our
information-geometric approach a relevant variable that will be useful for
comparison of the two different degrees of instability characterizing the
two ED models is the relative entropy-like quantity defined as,%
\begin{equation}
S_{1}\overset{\text{def}}{=}\ln \left( \Delta V_{1}\right) \text{.}
\end{equation}%
Substituting $\left( 50\right) $ in $\left( 51\right) $ and considering the
asymptotic limit $\tau \rightarrow \infty $, we obtain%
\begin{equation}
S_{1}\approx \alpha \tau \text{.}
\end{equation}%
The entropy-like quantity $S_{1}$ in $\left( 52\right) $ may be interpreted
as the asymptotic limit of the natural logarithm of a statistical weight $%
\left\langle \Delta V_{\mathcal{M}_{s_{1}}}\right\rangle _{\tau }$ defined
on $\mathcal{M}_{s_{1}}$. Equation $\left( 52\right) $ is the
information-geometric analog of the Zurek-Paz chaos criterion.

\subsubsection{The Jacobi Vector Field $J_{\mathcal{M}_{s_{1}}}$}

We study the behavior of the one-parameter family of neighboring geodesics $%
\mathcal{F}_{G_{\mathcal{M}_{s_{1}}}}\left( \alpha \right) \equiv \left\{
\theta _{\mathcal{M}_{s_{1}}}^{\mu }\left( \tau \text{; }\alpha \right)
\right\} _{\alpha \in 
\mathbb{R}
^{+}}^{\mu =1\text{, }2\text{, }3}$ where,%
\begin{eqnarray}
\theta ^{1}\left( \tau \text{; }\alpha \right) &=&\mu _{1}\left( \tau \text{%
; }\alpha \right) =Ae^{\alpha \tau }\text{, }\theta ^{2}\left( \tau \text{; }%
\alpha \right) =\mu _{2}\left( \tau \text{; }\alpha \right) =\frac{A^{2}}{%
2\alpha }\frac{1}{e^{-2\alpha \tau }+\frac{A^{2}}{8\alpha ^{2}}}\text{,} 
\notag \\
&& \\
\text{ }\theta ^{3}\left( \tau \text{; }\alpha \right) &=&\sigma _{2}\left(
\tau \text{; }\alpha \right) =A\frac{e^{-\alpha \tau }}{e^{-2\alpha \tau }+%
\frac{A^{2}}{8\alpha ^{2}}}\text{.}  \notag
\end{eqnarray}%
The relative geodesic spread is characterized by the Jacobi equation $\left[
16\text{, }17\right] $,%
\begin{equation}
\frac{D^{2}\left( \delta \theta ^{i}\right) }{D\tau ^{2}}+R_{kml}^{i}\frac{%
\partial \theta ^{k}}{\partial \tau }\frac{\partial \theta ^{l}}{\partial
\tau }\delta \theta ^{m}=0
\end{equation}%
where $i=1$, $2$, $3$ and,%
\begin{equation}
\delta \theta ^{i}\equiv \delta _{\alpha }\theta ^{i}\overset{\text{def}}{=}%
\left( \frac{\partial \theta ^{i}\left( \tau \text{; }\alpha \right) }{%
\partial \alpha }\right) _{\tau }\delta \alpha \text{.}
\end{equation}%
Equation $(54)$ forms a system of three coupled ordinary differential
equations \textit{linear} in the components of the deviation vector field $%
(55)$ but\textit{\ nonlinear} in derivatives of the metric $\left( 8\right) $%
. It describes the linearized geodesic flow: the linearization ignores the
relative velocity of the geodesics. When the geodesics are neighboring but
their relative velocity is arbitrary, the corresponding geodesic deviation
equation is the so-called generalized Jacobi equation $\left[ 18\right] $.
The nonlinearity is due to the existence of velocity-dependent terms in the
system.

Neighboring geodesics accelerate relative to each other with a rate directly
measured by the curvature tensor $R_{\alpha \beta \gamma \delta }$.
Multiplying both sides of $\left( 54\right) $ by $g_{ij}$ and using the
standard symmetry properties of the Riemann curvature tensor, the geodesic
deviation equation becomes,%
\begin{equation}
g_{ji}\frac{D^{2}\left( \delta \theta ^{i}\right) }{D\tau ^{2}}+R_{lmkj}%
\frac{\partial \theta ^{k}}{\partial \tau }\frac{\partial \theta ^{l}}{%
\partial \tau }\delta \theta ^{m}=0\text{.}
\end{equation}%
Recall that the covariant derivative $\frac{D^{2}\left( \delta \theta ^{\mu
}\right) }{D\tau ^{2}}$ in $\left( 54\right) $ is defined as,%
\begin{eqnarray}
\frac{D^{2}\delta ^{\mu }}{D\tau ^{2}} &=&\frac{d^{2}\delta ^{\mu }}{d\tau
^{2}}+2\Gamma _{\alpha \beta }^{\mu }\frac{d\delta ^{\alpha }}{d\tau }\frac{%
d\theta ^{\beta }}{d\tau }+\Gamma _{\alpha \beta }^{\mu }\delta ^{\alpha }%
\frac{d^{2}\theta ^{\beta }}{d\tau ^{2}}+\Gamma _{\alpha \beta ,\nu }^{\mu }%
\frac{d\theta ^{\nu }}{d\tau }\frac{d\theta ^{\beta }}{d\tau }\delta
^{\alpha }+  \notag \\
&&+\Gamma _{\alpha \beta }^{\mu }\Gamma _{\rho \sigma }^{\alpha }\frac{%
d\theta ^{\sigma }}{d\tau }\frac{d\theta ^{\beta }}{d\tau }\delta ^{\rho }
\end{eqnarray}%
and that the only non-vanishing Riemann tensor component is $R_{2323}=-\frac{%
1}{\sigma _{1}^{4}}$. Therefore, the three differential equations for the
geodesic deviation are,%
\begin{equation}
\frac{d^{2}\left( \delta \theta ^{1}\right) }{d\tau ^{2}}+2\Gamma _{11}^{1}%
\frac{d\theta ^{1}}{d\tau }\frac{d\left( \delta \theta ^{1}\right) }{d\tau }%
+\partial _{1}\Gamma _{11}^{1}\left( \frac{d\theta ^{1}}{d\tau }\right)
^{2}\delta \theta ^{1}=0\text{,}
\end{equation}%
\begin{eqnarray}
&&\frac{d^{2}\left( \delta \theta ^{2}\right) }{d\tau ^{2}}+2\left[ \Gamma
_{23}^{2}\frac{d\theta ^{3}}{d\tau }\frac{d\left( \delta \theta ^{2}\right) 
}{d\tau }+\Gamma _{32}^{2}\frac{d\theta ^{2}}{d\tau }\frac{d\left( \delta
\theta ^{3}\right) }{d\tau }\right] +\partial _{3}\Gamma _{23}^{2}\left( 
\frac{d\theta ^{3}}{d\tau }\right) ^{2}\delta \theta ^{2}+\Gamma
_{32}^{2}\Gamma _{33}^{3}\left( \frac{d\theta ^{3}}{d\tau }\right)
^{2}\delta \theta ^{2}  \notag \\
&=&\frac{1}{g_{22}}R_{2323}\frac{d\theta ^{2}}{d\tau }\frac{d\theta ^{3}}{%
d\tau }\delta \theta ^{3}-\frac{1}{g_{22}}R_{2323}\left( \frac{d\theta ^{3}}{%
d\tau }\right) ^{2}\delta \theta ^{2}\text{,}
\end{eqnarray}%
\begin{eqnarray}
&&\frac{d^{2}\left( \delta \theta ^{3}\right) }{d\tau ^{2}}+2\left[ \Gamma
_{22}^{3}\frac{d\theta ^{2}}{d\tau }\frac{d\left( \delta \theta ^{2}\right) 
}{d\tau }+\Gamma _{33}^{3}\frac{d\theta ^{3}}{d\tau }\frac{d\left( \delta
\theta ^{3}\right) }{d\tau }\right] +\partial _{3}\Gamma _{33}^{3}\left( 
\frac{d\theta ^{3}}{d\tau }\right) ^{2}\delta \theta ^{3}+\Gamma
_{22}^{3}\Gamma _{23}^{2}\frac{d\theta ^{3}}{d\tau }\frac{d\theta ^{2}}{%
d\tau }\delta \theta ^{2}  \notag \\
&=&\frac{1}{g_{33}}R_{2323}\frac{d\theta ^{2}}{d\tau }\frac{d\theta ^{3}}{%
d\tau }\delta \theta ^{2}-\frac{1}{g_{33}}R_{2323}\left( \frac{d\theta ^{2}}{%
d\tau }\right) ^{2}\delta \theta ^{3}\text{.}
\end{eqnarray}%
Substituting $\left( 14\right) $, $\left( 16\right) $ and $\left( 53\right) $
in equations $\left( 58\right) $, $\left( 59\right) $ and $\left( 60\right) $
and considering the asymptotic limit $\tau \rightarrow \infty $, the
geodesic deviation equations become,%
\begin{equation}
\frac{d^{2}\left( \delta \theta ^{1}\right) }{d\tau ^{2}}+2\alpha \frac{%
d\left( \delta \theta ^{1}\right) }{d\tau }+\alpha ^{2}\delta \theta ^{1}=0%
\text{,}
\end{equation}%
\begin{equation}
\frac{d^{2}\left( \delta \theta ^{2}\right) }{d\tau ^{2}}+2\alpha \frac{%
d\left( \delta \theta ^{2}\right) }{d\tau }+\frac{16\alpha ^{2}}{A}%
e^{-\alpha \tau }\frac{d\left( \delta \theta ^{3}\right) }{d\tau }+\left(
\alpha ^{2}-\frac{8\alpha ^{3}}{A}e^{-\alpha \tau }\right) \delta \theta
^{3}=0\text{,}
\end{equation}%
\begin{equation}
\frac{d^{2}\left( \delta \theta ^{3}\right) }{d\tau ^{2}}+2\alpha \frac{%
d\left( \delta \theta ^{3}\right) }{d\tau }+\left( \alpha ^{2}-\frac{%
32\alpha ^{4}}{A^{2}}e^{-2\alpha \tau }\right) \delta \theta ^{3}-\frac{%
8\alpha ^{2}}{A}e^{-\alpha \tau }\frac{d\left( \delta \theta ^{2}\right) }{%
d\tau }-\frac{8\alpha ^{3}}{A}e^{-\alpha \tau }\delta \theta ^{2}=0\text{.}
\end{equation}%
Neglecting the exponentially decaying terms in $\delta \theta ^{3}$ in $%
\left( 62\right) $ and $\left( 63\right) $ and assuming that,%
\begin{equation}
\underset{\tau \rightarrow \infty }{\lim }\left( \frac{16\alpha ^{2}}{A}%
e^{-\alpha \tau }\frac{d\left( \delta \theta ^{3}\right) }{d\tau }\right) =0%
\text{, }\underset{\tau \rightarrow \infty }{\lim }\left( \frac{8\alpha ^{2}%
}{A}e^{-\alpha \tau }\frac{d\left( \delta \theta ^{2}\right) }{d\tau }%
\right) =0\text{,}\underset{\tau \rightarrow \infty }{\lim }\left( \frac{%
8\alpha ^{3}}{A}e^{-\alpha \tau }\delta \theta ^{2}\right) =0\text{, }
\end{equation}%
the geodesic deviation equations finally become,%
\begin{eqnarray}
\frac{d^{2}\left( \delta \theta ^{1}\right) }{d\tau ^{2}}+2\alpha \frac{%
d\left( \delta \theta ^{1}\right) }{d\tau }+\alpha ^{2}\delta \theta ^{1}
&=&0\text{, }\frac{d^{2}\left( \delta \theta ^{2}\right) }{d\tau ^{2}}%
+2\alpha \frac{d\left( \delta \theta ^{2}\right) }{d\tau }+\alpha ^{2}\delta
\theta ^{3}=0\text{,}  \notag \\
\frac{d^{2}\left( \delta \theta ^{3}\right) }{d\tau ^{2}}+2\alpha \frac{%
d\left( \delta \theta ^{3}\right) }{d\tau }+\alpha ^{2}\delta \theta ^{3}
&=&0\text{.}
\end{eqnarray}%
Note that in order to prove that our assumptions in $\left( 64\right) $ are
correct, we will check \textit{a posteriori }its consistency. Integrating
the system of differential equations $\left( 65\right) $, we obtain%
\begin{eqnarray}
\delta \mu _{1}\left( \tau \right) &=&\left( a_{1}+a_{2}\tau \right)
e^{-\alpha \tau }\text{, }\delta \mu _{2}\left( \tau \right) =\left(
a_{3}+a_{4}\tau \right) e^{-\alpha \tau }-\frac{1}{2\alpha }a_{5}e^{-2\alpha
\tau }+a_{6}\text{,} \\
\delta \sigma _{2}\left( \tau \right) &=&\left( a_{3}+a_{4}\tau \right)
e^{-\alpha \tau }\text{,}  \notag
\end{eqnarray}%
where $a_{i}$, $i=1$,..., $6$ are integration constants. Note that
conditions $\left( 64\right) $ are satisfied and therefore our assumption
are compatible with the solutions obtained. Finally, consider the vector
field components $J^{k}\equiv \delta \theta ^{k}$ defined in $\left(
55\right) $ and its magnitude $J$, \ \ 
\begin{equation}
J^{2}=J^{i}J_{i}=g_{ij}J^{i}J^{j}\text{.}
\end{equation}%
The magnitude $J$ is called the Jacobi field intensity. In our case $\left(
67\right) $ becomes,%
\begin{equation}
J_{\mathcal{M}_{s_{1}}}^{2}=\frac{1}{\mu _{1}^{2}}\left( \delta \mu
_{1}\right) ^{2}+\frac{1}{\sigma _{2}^{2}}\left( \delta \mu _{2}\right) ^{2}+%
\frac{2}{\sigma _{2}^{2}}\left( \delta \sigma _{2}\right) ^{2}\text{.}
\end{equation}%
Substituting $\left( 53\right) $ and $\left( 66\right) $ in $\left(
68\right) $, and keeping the leading term in the asymptotic expansion in $J_{%
\mathcal{M}_{s_{1}}}^{2}$, we obtain%
\begin{equation}
J_{\mathcal{M}_{s_{1}}}\approx C_{\mathcal{M}_{s_{1}}}e^{\alpha \tau }\text{,%
}
\end{equation}%
where the constant coefficient $C_{\mathcal{M}_{s_{1}}}=$ $\frac{Aa_{6}^{3}}{%
2\sqrt{2}\alpha }$ encodes information about initial conditions and depends
on the model parameter $\alpha $. We conclude that the geodesic spread on $%
\mathcal{M}_{s_{1}}$ is described by means of an \textit{exponentially} 
\textit{divergent} Jacobi vector field intensity $J_{\mathcal{M}_{s_{1}}}$.
It is known that classical chaotic systems exhibit exponential sensitivity
to initial conditions. This characterization, quantified in terms of
Lyapunov exponents, is an important ingredient in any conventional
definition of classical chaos. In our approach, the quantity $\lambda
_{J}\approx \frac{1}{\tau }\underset{\tau \rightarrow \infty }{\lim }\ln %
\left[ \frac{\left\vert J\left( \tau \right) \right\vert }{\left\vert
J\left( 0\right) \right\vert }\right] $ with $J$ given in $\left( 69\right) $
would play the role of the conventional Lyapunov exponents.

\subsection{Instability in RED2}

In this subsection, the instability of the geodesics on $\mathcal{M}_{s_{2}}$
is studied. We proceed as in subsection $\left( 4.1\right) $.

\subsubsection{The Geodesic Length $\Theta _{\mathcal{M}_{s_{2}}}$}

Consider the one-parameter family of geodesics $\mathcal{F}_{G_{\mathcal{M}%
_{s_{2}}}}\left( \alpha \right) \equiv \left\{ \theta _{\mathcal{M}%
_{s_{2}}}^{\mu }\left( \tau \text{; }\alpha \right) \right\} _{\alpha \in 
\mathbb{R}
^{+}}^{\mu =1,2,3,4}$ where $\theta _{\mathcal{M}_{s_{2}}}^{\mu }$are
solutions of $\left( 39\right) $. The length of geodesics in $\mathcal{F}%
_{G_{\mathcal{M}_{s_{2}}}}\left( \alpha \right) $ is defined as,%
\begin{equation}
\Theta _{\mathcal{M}_{s_{2}}}\left( \tau \text{; }\alpha \right) \overset{%
\text{def}}{=}\dint \left( g_{ij}d\theta ^{i}d\theta ^{j}\right) ^{\frac{1}{2%
}}=\dint\limits_{0}^{\tau }\left[ \frac{1}{\sigma _{1}^{2}}\left( \frac{d\mu
_{1}}{d\tau ^{\prime }}\right) ^{2}+\frac{2}{\sigma _{1}^{2}}\left( \frac{%
d\sigma _{1}}{d\tau ^{\prime }}\right) ^{2}+\frac{1}{\sigma _{2}^{2}}\left( 
\frac{d\mu _{2}}{d\tau ^{\prime }}\right) ^{2}+\frac{2}{\sigma _{2}^{2}}%
\left( \frac{d\sigma _{2}}{d\tau ^{\prime }}\right) ^{2}\right] ^{\frac{1}{2}%
}d\tau ^{\prime }\text{.}
\end{equation}%
Substituting $\left( 40\right) $ in $\left( 70\right) $ and considering the
asymptotic limit of $\Theta _{\mathcal{M}_{s_{2}}}\left( \tau \text{; }%
\alpha \right) $ when $\tau \rightarrow \infty $, we obtain,%
\begin{equation}
\Theta _{\mathcal{M}_{s_{2}}}\left( \tau \rightarrow \infty \text{; }\alpha
\right) \equiv \Theta _{2}\left( \tau \text{; }\alpha \right) \approx
2\alpha \tau \text{.}
\end{equation}%
In order to investigate the asymptotic behavior of two neighboring geodesics
labelled by the parameters $\alpha $ and $\alpha +\delta \alpha $, we
consider the following difference,%
\begin{equation}
\Delta \Theta _{2}\equiv \left\vert \Theta _{2}\left( \tau \text{; }\alpha
+\delta \alpha \right) -\Theta _{2}\left( \tau \text{; }\alpha \right)
\right\vert =2\left\vert \delta \alpha \right\vert \tau \text{.}
\end{equation}%
It is clear that $\Delta \Theta _{2}$ diverges, that is the lengths of two
neighboring geodesics\ with slightly different parameters $\alpha $ and $%
\alpha +\delta \alpha $ differ in a significant way as the evolution
parameter $\tau \rightarrow \infty $. This hints at the onset of instability
of the hyperbolic trajectories on $\mathcal{M}_{s_{2}}$.

\subsubsection{The Statistical Volume Elements $V_{\mathcal{M}_{s_{2}}}$}

The instability of ED2 can be explored by studying the behavior of the
one-parameter family of statistical volume elements $\mathcal{F}_{V_{%
\mathcal{M}_{s_{2}}}}\left( \alpha \right) \equiv \left\{ V_{\mathcal{M}%
_{s_{2}}}\left( \tau \text{; }\alpha \right) \right\} _{\alpha }$. Recall
that $\mathcal{M}_{s_{2}}$ is the space of probability distributions $p^{(%
\text{tot})}\left( \vec{x}|\vec{\theta}\right) $ labeled by parameters $%
\theta _{1}^{\left( 1\right) }$, $\theta _{2}^{\left( 1\right) }$, $\theta
_{1}^{\left( 2\right) }$, $\theta _{2}^{\left( 2\right) }$. These parameters
are the coordinates of the point $p^{(\text{tot})}$, and in these
coordinates a $4D$ infinitesimal volume element $dV_{\mathcal{M}_{s_{2}}}$
reads,%
\begin{equation}
dV_{\mathcal{M}_{s_{2}}}=\sqrt{g}d\theta _{1}^{\left( 1\right) }d\theta
_{2}^{\left( 1\right) }d\theta _{1}^{\left( 2\right) }d\theta _{2}^{\left(
2\right) }\equiv \sqrt{g}d\mu _{1}d\sigma _{1}d\mu _{2}d\sigma _{2}\text{,}
\end{equation}%
where in the ED2 model here presented, $g=|\det \left( g_{ij}\right) _{%
\mathcal{M}_{s_{2}}}|=\frac{4}{\sigma _{1}^{4}\sigma _{2}^{4}}$. Hence, the
infinitesimal volume element $dV_{\mathcal{M}_{s_{2}}}$ is given by,%
\begin{equation}
dV_{\mathcal{M}_{s_{2}}}=\frac{2}{\sigma _{1}^{2}\sigma _{2}^{2}}d\mu
_{1}d\sigma _{1}d\mu _{2}d\sigma _{2}\text{.}
\end{equation}%
The volume of an extended region of $\mathcal{M}_{s_{2}}$ is defined by,%
\begin{equation}
\Delta V_{\mathcal{M}_{s_{2}}}\equiv V_{\mathcal{M}_{s_{2}}}\left( \tau
\right) -V_{\mathcal{M}_{s_{2}}}\left( 0\right) \overset{\text{def}}{=}%
\dint\limits_{V_{\mathcal{M}_{s_{2}}}\left( 0\right) }^{V_{\mathcal{M}%
_{s_{2}}}\left( \tau \right) }dV_{\mathcal{M}_{s_{2}}}=\dint\limits_{\mu
_{1}\left( 0\right) }^{\mu _{1}\left( \tau \right) }\dint\limits_{\sigma
_{1}\left( 0\right) }^{\sigma _{1}\left( \tau \right) }\text{ }%
\dint\limits_{\mu _{2}\left( 0\right) }^{\mu _{2}\left( \tau \right)
}\dint\limits_{\sigma _{2}\left( 0\right) }^{\sigma _{2}\left( \tau \right) }%
\frac{2}{\sigma _{1}^{2}\sigma _{2}^{2}}d\mu _{1}d\sigma _{1}d\mu
_{2}d\sigma _{2}\text{,}
\end{equation}%
Integrating $\left( 75\right) $ and using $\left( 40\right) $, we obtain%
\begin{equation}
\Delta V_{\mathcal{M}_{s_{2}}}=\frac{A^{2}}{2\alpha ^{2}}e^{2\alpha \tau }-%
\frac{A^{2}}{2\alpha ^{2}}\text{.}
\end{equation}
The average volume on $\mathcal{M}_{s_{2}}$ is $\left\langle \Delta V_{%
\mathcal{M}_{s_{2}}}\right\rangle _{\tau }$, 
\begin{equation}
\left\langle \Delta V_{\mathcal{M}_{s_{2}}}\right\rangle _{\tau }\overset{%
\text{def}}{=}\frac{1}{\tau }\dint\limits_{0}^{\tau }\Delta V_{\mathcal{M}%
_{s_{2}}}\left( \tau ^{\prime }\text{; }\alpha \right) d\tau ^{\prime }=%
\frac{A^{2}}{4\alpha ^{3}}\frac{e^{2\alpha \tau }}{\tau }-\frac{A^{2}}{%
2\alpha ^{2}}
\end{equation}%
For convenience, let us rename $\left\langle \Delta V_{\mathcal{M}%
_{s_{2}}}\right\rangle _{\tau }\equiv \Delta V_{2}$. Therefore, the
asymptotic expansion of $\Delta V_{2}$ for $\tau $ $\rightarrow \infty $
reads,%
\begin{equation}
\Delta V_{2}\approx \frac{A^{2}}{4\alpha ^{3}}\frac{e^{2\alpha \tau }}{\tau }%
\text{.}
\end{equation}%
In analogy to $\left( 51\right) $ we introduce,%
\begin{equation}
S_{2}\overset{\text{def}}{=}\ln \left( \Delta V_{2}\right) \text{.}
\end{equation}%
Substituting $\left( 78\right) $ in $\left( 79\right) $ and considering its
asymptotic limit, we obtain%
\begin{equation}
S_{2}\approx 2\alpha \tau \text{.}
\end{equation}

\subsubsection{The Jacobi Vector Field $J_{\mathcal{M}_{s_{2}}}$}

We proceed as in $\left( 4.1.3\right) $. Study the behavior of the
one-parameter $\left( \alpha \right) $ family of neighboring geodesics on $%
\mathcal{M}_{s_{2}}$, $\left\{ \theta ^{i}\left( \tau \text{; }\alpha
\right) \right\} _{i=1\text{, }2\text{, }3\text{, }4}$ with%
\begin{equation}
\theta ^{3}\left( \tau ;\alpha \right) \equiv \theta ^{1}\left( \tau ;\alpha
\right) =\mu _{1}\left( \tau ;\alpha \right) =\frac{A^{2}}{2\alpha }\frac{1}{%
e^{-2\alpha \tau }+\frac{A^{2}}{8\alpha ^{2}}}\text{,}
\end{equation}%
\begin{equation}
\theta ^{4}\left( \tau ;\alpha \right) \equiv \theta ^{2}\left( \tau ;\alpha
\right) =A\frac{e^{-\alpha \tau }}{e^{-2\alpha \tau }+\frac{A^{2}}{8\alpha
^{2}}}\text{.}
\end{equation}%
Note that because we will compare the two Jacobi fields $J_{\mathcal{M}%
_{s_{1}}}$ on $\mathcal{M}_{s_{1}}$ and $J_{\mathcal{M}_{s_{2}}}$ on $%
\mathcal{M}_{s_{2}}$, we assume the same initial conditions as considered in 
$\left( 4.1.3\right) $. Recall that the non-vanishing Riemann tensor
components are $R_{1212}=-\frac{1}{\sigma _{1}^{4}}$ and $R_{3434}=-\frac{1}{%
\sigma _{2}^{4}}$ given in $\left( 27\right) $. Therefore two of the four
differential equations describing the geodesic spread are,%
\begin{eqnarray}
&&\frac{d^{2}\left( \delta \theta ^{1}\right) }{d\tau ^{2}}+2\left[ \Gamma
_{12}^{1}\frac{d\theta ^{2}}{d\tau }\frac{d\left( \delta \theta ^{1}\right) 
}{d\tau }+\Gamma _{21}^{1}\frac{d\theta ^{1}}{d\tau }\frac{d\left( \delta
\theta ^{2}\right) }{d\tau }\right] +\partial _{2}\Gamma _{12}^{1}\left( 
\frac{d\theta ^{2}}{d\tau }\right) ^{2}\delta \theta ^{1}+\Gamma
_{21}^{1}\Gamma _{22}^{2}\left( \frac{d\theta ^{2}}{d\tau }\right)
^{2}\delta \theta ^{1}  \notag \\
&=&\frac{1}{g_{11}}R_{1212}\frac{d\theta ^{1}}{d\tau }\frac{d\theta ^{2}}{%
d\tau }\delta \theta ^{2}-\frac{1}{g_{11}}R_{1212}\left( \frac{d\theta ^{2}}{%
d\tau }\right) ^{2}\delta \theta ^{1}\text{,}
\end{eqnarray}%
\begin{eqnarray}
&&\frac{d^{2}\left( \delta \theta ^{2}\right) }{d\tau ^{2}}+2\left[ \Gamma
_{11}^{2}\frac{d\theta ^{1}}{d\tau }\frac{d\left( \delta \theta ^{1}\right) 
}{d\tau }+\Gamma _{22}^{2}\frac{d\theta ^{2}}{d\tau }\frac{d\left( \delta
\theta ^{2}\right) }{d\tau }\right] +\partial _{2}\Gamma _{22}^{2}\left( 
\frac{d\theta ^{2}}{d\tau }\right) ^{2}\delta \theta ^{2}+\Gamma
_{11}^{2}\Gamma _{12}^{1}\frac{d\theta ^{2}}{d\tau }\frac{d\theta ^{1}}{%
d\tau }\delta \theta ^{1}  \notag \\
&=&\frac{1}{g_{22}}R_{1212}\frac{d\theta ^{1}}{d\tau }\frac{d\theta ^{2}}{%
d\tau }\delta \theta ^{1}-\frac{1}{g_{22}}R_{1212}\left( \frac{d\theta ^{1}}{%
d\tau }\right) ^{2}\delta \theta ^{2}\text{.}
\end{eqnarray}%
The other two equations can be obtained from $\left( 83\right) $ and $\left(
84\right) $ substituting the index $1$ with $3$ and $2$ with $4$. Thus, we
will limit our considerations just to the above two equations. Using
equations $\left( 25\right) $, $\left( 27\right) $, $\left( 81\right) $ and $%
\left( 82\right) $ in $\left( 83\right) $ and $\left( 84\right) $ and
considering the asymptotic limit $\tau \rightarrow \infty $, the two
equations of geodesic deviation become,%
\begin{equation}
\frac{d^{2}\left( \delta \theta ^{1}\right) }{d\tau ^{2}}+2\alpha \frac{%
d\left( \delta \theta ^{1}\right) }{d\tau }+\frac{16\alpha ^{2}}{A}%
e^{-\alpha \tau }\frac{d\left( \delta \theta ^{2}\right) }{d\tau }+\left(
\alpha ^{2}-\frac{8\alpha ^{3}}{A}e^{-\alpha \tau }\right) \delta \theta
^{2}=0\text{,}
\end{equation}%
\begin{equation}
\frac{d^{2}\left( \delta \theta ^{2}\right) }{d\tau ^{2}}+2\alpha \frac{%
d\left( \delta \theta ^{2}\right) }{d\tau }+\left( \alpha ^{2}-\frac{%
32\alpha ^{4}}{A^{2}}e^{-2\alpha \tau }\right) \delta \theta ^{2}-\frac{%
8\alpha ^{2}}{A}e^{-\alpha \tau }\frac{d\left( \delta \theta ^{1}\right) }{%
d\tau }-\frac{8\alpha ^{3}}{A}e^{-\alpha \tau }\delta \theta ^{1}=0.
\end{equation}%
Neglecting the exponentially decaying terms in $\delta \theta ^{2}$ in $%
\left( 85\right) $ and $\left( 86\right) $ and assuming%
\begin{equation}
\underset{\tau \rightarrow \infty }{\lim }\left( \frac{16\alpha ^{2}}{A}%
e^{-\alpha \tau }\frac{d\left( \delta \theta ^{2}\right) }{d\tau }\right) =0%
\text{,}\underset{\tau \rightarrow \infty }{\lim }\left( \frac{8\alpha ^{2}}{%
A}e^{-\alpha \tau }\frac{d\left( \delta \theta ^{1}\right) }{d\tau }\right)
=0\text{,}\underset{\tau \rightarrow \infty }{\lim }\left( \frac{8\alpha ^{3}%
}{A}e^{-\alpha \tau }\delta \theta ^{1}\right) =0
\end{equation}%
the geodesic deviation equations in $\left( 85\right) $ and $\left( \text{ }%
86\right) $ become,%
\begin{equation}
\frac{d^{2}\left( \delta \theta ^{1}\right) }{d\tau ^{2}}+2\alpha \frac{%
d\left( \delta \theta ^{1}\right) }{d\tau }+\alpha ^{2}\delta \theta ^{2}=0%
\text{, }\frac{d^{2}\left( \delta \theta ^{2}\right) }{d\tau ^{2}}+2\alpha 
\frac{d\left( \delta \theta ^{2}\right) }{d\tau }+\alpha ^{2}\delta \theta
^{2}=0\text{.}
\end{equation}%
The consistency of the assumptions in $\left( 87\right) $ will be checked 
\textit{a posteriori} after integrating equations in $\left( 88\right) $. It
follows that the geodesics spread on $\mathcal{M}_{s_{2}}$ is described by
the temporal evolution of the following deviation vector components,%
\begin{eqnarray}
\delta \mu _{1}\left( \tau \right) &=&\left( a_{1}+a_{2}\tau \right)
e^{-\alpha \tau }-\frac{1}{2\alpha }a_{3}e^{-2\alpha \tau }+a_{4}\text{, }%
\delta \sigma _{1}\left( \tau \right) =\left( a_{1}+a_{2}\tau \right)
e^{-\alpha \tau } \\
\text{ }\delta \mu _{2}\left( \tau \right) &=&\left( a_{5}+a_{6}\tau \right)
e^{-\alpha \tau }-\frac{1}{2\alpha }a_{7}e^{-2\alpha \tau }+a_{8}\text{, }%
\delta \sigma _{2}\left( \tau \right) =\left( a_{5}+a_{6}\tau \right)
e^{-\alpha \tau }  \notag
\end{eqnarray}%
where $a_{i}$, $i=1$,..., $8$ are integration constants. Note that $a_{4}$
and $a_{8}$ in $\left( 89\right) $ equal $a_{6}$ in $\left( 66\right) $.
Furthermore, note that these solutions above are compatible with the
assumptions in $\left( 87\right) $. Finally, consider the Jacobi vector
field intensity $J_{\mathcal{M}_{s_{2}}}$ on $\mathcal{M}_{s_{2}}$,%
\begin{equation}
J_{\mathcal{M}_{s_{2}}}^{2}=\frac{1}{\sigma _{1}^{2}}\left( \delta \mu
_{1}\right) ^{2}+\frac{2}{\sigma _{1}^{2}}\left( \delta \sigma _{1}\right)
^{2}+\frac{1}{\sigma _{2}^{2}}\left( \delta \mu _{2}\right) ^{2}+\frac{2}{%
\sigma _{2}^{2}}\left( \delta \sigma _{2}\right) ^{2}\text{.}
\end{equation}%
Substituting $\left( 81\right) $, $\left( 82\right) $ and $\left( 89\right) $
in $\left( 90\right) $ and keeping the leading term in the asymptotic
expansion in $J_{\mathcal{M}_{s_{2}}}^{2}$, we obtain%
\begin{equation}
J_{\mathcal{M}_{s_{2}}}\approx C_{\mathcal{M}_{s_{2}}}e^{\alpha \tau }\text{.%
}
\end{equation}%
where the constant coefficient $C_{\mathcal{M}_{s_{2}}}=$ $\frac{Aa_{6}^{3}}{%
\sqrt{2}\alpha }\equiv 2C_{\mathcal{M}_{s_{1}}}$ encodes information about
initial conditions and it depends on the model parameter $\alpha $. We
conclude that the geodesic spread on $\mathcal{M}_{s_{2}}$ is described by
means of an \textit{exponentially} divergent Jacobi vector field intensity $%
J_{\mathcal{M}_{s_{2}}}$.

\section{Statistical Curvature, Jacobi Field Intensity and Entropy-like
Quantities}

It is known that statistical manifolds allow differential geometric methods
to be applied to problems in mathematical statistics, information theory and
in stochastic processes. For instance, an important class of statistical
manifolds is that arising from the so-called exponential family $\left[ 8%
\right] $ and one particular family is that of gamma probability
distributions. These distributions have been shown $\left[ 19\right] $ to
have important uniqueness properties for near-random stochastic processes.
In this paper, two statistical manifolds of negative curvature $\mathcal{M}%
_{s_{1}}$ and $\mathcal{M}_{s_{2}}$ have been considered. They are
representations of smooth families of probability distributions
(exponentials and Gaussians for $\mathcal{M}_{s_{1}}$, Gaussians for $%
\mathcal{M}_{s_{2}}$). They represent the "arena" where the entropic
dynamics takes place. The instability of the trajectories of the ED1 and ED2
on $\mathcal{M}_{s_{1}}$ and $\mathcal{M}_{s_{2}}$ respectively, has been
studied using the statistical weight $\left\langle \Delta V_{\mathcal{M}%
_{s}}\right\rangle _{\tau }$ defined on the curved manifold $\mathcal{M}_{s}$
and the Jacobi vector field intensity $J_{\mathcal{M}_{s}}$. Does our
analysis lead to any possible further understanding of the role of
statistical curvature in physics and statistics? We argue that it does.

The role of curvature in physics is fairly well understood. It encodes
information about the field strengths for all the four fundamental
interactions in nature. The curvature plays a key role in the Riemannian
geometric approach to chaos $\left[ 20\right] $. In this approach, the study
of the Hamiltonian dynamics is reduced to the investigation of geometrical
properties of the manifold on which geodesic flow is induced. For instance,
the stability or local instability of geodesic flows depends on the
sectional curvature properties of the suitable defined metric manifold. The
sectional curvature brings the same qualitative and quantitative information
that is provided by the Lyapunov exponents in the conventional approach.
Furthermore, the integrability of the system is connected with existence of
Killing vectors on the manifold. However, a rigorous relation among
curvature, Lyapunov exponents and Kolmogorov-Sinay entropy $\left[ 21\right] 
$ is still under investigation. In addition, there does not exist a well
defined unifying characterization of chaos in classical and quantum physics $%
\left[ 22\right] $ due to fundamental differences between the two theories.
In addition, the role of curvature in statistical inference is even less
understood. The meaning of statistical curvature for a one-parameter model
in inference theory was introduced in $\left[ 23\right] $. Curvature served
as an important tool in the asymptotic theory of statistical estimation. The
higher the scalar curvature at a given point on the manifold, the more
difficult it is to do estimation at that point $\left[ 24\right] $.

Recall that the entropy-like quantity $S$ is the asymptotic limit of the
natural logarithm of the average portion of the statistical volume $%
\left\langle \Delta V_{\mathcal{M}_{s}}\right\rangle _{\tau }\ $associated
to the evolution of the geodesics on $\mathcal{M}_{s}$ . Considering
equations $\left( 52\right) $ and $\left( 80\right) $, we obtain,%
\begin{equation}
S_{2}\approx 2S_{1}\text{.}
\end{equation}%
Furthermore, the relationship between the statistical curvatures on the
curved manifolds $\mathcal{M}_{s_{1}}$ and $\mathcal{M}_{s_{2}}$ is, 
\begin{equation}
R_{\mathcal{M}_{s_{2}}}=2R_{\mathcal{M}_{s_{1}}}\text{.}
\end{equation}
In view of $\left( 92\right) $ and $\left( 93\right) $, it follows that
there is a direct proportionality between the curvature $R_{\mathcal{M}_{s}}$
and the asymptotic expression for the entropy-like quantity $S$
characterizing the ED on manifolds $\mathcal{M}_{s_{i}}$ with $i=1$, $2$,
namely%
\begin{equation}
\frac{R_{\mathcal{M}_{s_{2}}}}{R_{\mathcal{M}_{s_{1}}}}=\frac{S_{2}}{S_{1}}%
\text{.}
\end{equation}%
Moreover, from $\left( 69\right) $ and $\left( 91\right) $, we obtain the
following relation,%
\begin{equation}
J_{\mathcal{M}_{s_{2}}}\approx 2J_{\mathcal{M}_{s_{1}}}\text{.}
\end{equation}%
The two manifolds $\mathcal{M}_{s_{1}}$ and $\mathcal{M}_{s_{2}}$ are
exponentially unstable and the intensity of Jacobi vector field $J_{\mathcal{%
M}_{s_{2}}}$ of manifold $\mathcal{M}_{s_{2}}$ with curvature $R_{\mathcal{M}%
_{s_{2}}}=-2$ is asymptotically twice the intensity of the Jacobi vector
field $J_{\mathcal{M}_{s_{1}}}$of manifold $\mathcal{M}_{s_{1}}$with
curvature $R=-1$. Considering $\left( 93\right) $ and $\left( 95\right) $,
we obtain 
\begin{equation}
\frac{R_{\mathcal{M}_{s_{2}}}}{R_{\mathcal{M}_{s_{1}}}}=\frac{J_{\mathcal{M}%
_{s_{2}}}}{J_{\mathcal{M}_{s_{1}}}}\text{.}
\end{equation}%
It seems there exists a direct proportionality between the curvature $R_{%
\mathcal{M}_{s}}$ and the intensity of the Jacobi field $J_{\mathcal{M}_{s}}$
characterizing the degree of chaoticity of a statistical manifold of
negative curvature $\mathcal{M}_{s}$. Finally, comparison of $\left(
94\right) $ and $\left( 96\right) $ leads to the formal link between
curvature, entropy and chaoticity:%
\begin{equation}
R\sim S\sim J\text{.}
\end{equation}%
Though several points need deeper understanding and analysis, we hope that
our work convincingly shows that this information-geometric approach may be
useful in providing a unifying criterion of chaos of both classical and
quantum varieties, thus deserving further research and developments.

\section{Conclusions}

Two entropic dynamical models have been considered: a $3D$ and $4D$
statistical manifold $\mathcal{M}_{s_{1}}$ and $\mathcal{M}_{s_{2}}$
respectively. These manifolds serve as the stage on which the entropic
dynamics takes place. In the former case, macro-coordinates on the manifold
are represented by the expectation values of microvariables associated with
Gaussian and exponential probability distributions. In the latter case,
macro-coordinates are expectation values of microvariables associated with
two Gaussians distributions. The geometric structure of $\mathcal{M}_{s_{1}}$
and $\mathcal{M}_{s_{2}}$ was studied in detail. It was shown that $\mathcal{%
M}_{s_{1}}$ is a curved manifold of constant negative curvature $-1$ while $%
\mathcal{M}_{s_{2}}$ has constant negative curvature $-2$. The geodesics of
the ED models are hyperbolic curves on $\mathcal{M}_{s_{i}}$ $\left( i=1%
\text{, }2\right) $. A study of the stability of geodesics on $\mathcal{M}%
_{s_{1}}$ and $\mathcal{M}_{s_{2}}$ was presented. The notion of statistical
volume elements was introduced to investigate the asymptotic behavior of a
one-parameter family of neighboring volumes $\mathcal{F}_{V_{\mathcal{M}%
_{s}}}\left( \alpha \right) \equiv \left\{ V_{\mathcal{M}_{s}}\left( \tau 
\text{; }\alpha \right) \right\} _{\alpha }$. An information-geometric
analog of the Zurek-Paz chaos criterion was presented. It was shown that the
behavior of geodesics is characterized by exponential instability that leads
to chaotic scenarios on the curved statistical manifolds. These conclusions
are supported by a study based on the geodesic deviation equations and on
the asymptotic behavior of the Jacobi vector field intensity $J_{\mathcal{M}%
_{s}}$ on $\mathcal{M}_{s_{1}}$ and $\mathcal{M}_{s_{2}}$. A Lyapunov
exponent analog similar to that appearing in the Riemannian geometric
approach was suggested as an indicator of chaos. On the basis of our
analysis a relationship among an entropy-like quantity, chaoticity and
curvature in the two models ED1 and ED2 is proposed, suggesting to interpret
the statistical curvature as a measure of the entropic dynamical chaoticity.

The implications of this work is twofold. Firstly, it helps understanding
possible future use of the statistical curvature in modelling real processes
by relating it to conventionally accepted quantities such as entropy and
chaos. On the other hand, it serves to cast what is already known in physics
regarding curvature in a new light as a consequence of its proposed link
with inference. Finally we remark that based on the results obtained from
the chosen ED models, it is not unreasonable to think that should the
correct variables describing the true degrees of freedom of a physical
system be identified, perhaps deeper insights into the foundations of models
of physics and reasoning (and their relationship to each other) may be
uncovered.

\noindent \textbf{Acknowledgements:} The authors are grateful to Dr. A. Caticha for useful discussions.

\end{document}